\documentstyle[12pt]{article}

\begin {document}

\begin {center}
{\large\bf Temperatures of Fragment Kinetic Energy Spectra\\[5ex]}

{\bf Wolfgang Bauer}\\[2ex]

\begin{it}
Institute for Nuclear Theory, University of Washington\\
Seattle, WA 98195, USA\\
{\rm and}\\
National Superconducting Cyclotron Laboratory and\\
Department of Physics and Astronomy, Michigan State University\\
East Lansing, MI 48824, USA\\[5ex]
\end{it}
\end {center}

\begin{abstract}
Multifragmentation reactions without large compression in the initial state
(proton-induced reactions, reverse-kinematics, projectile fragmentation)
are examined, and it is
shown that the high temperatures obtained from fragment kinetic energy
spectra and lower temperatures obtained from observables such as level
population or isotope ratios can be understood in a common framework.
\end{abstract}

\newpage

The phenomenon of nuclear multifragmentation, the decay of highly excited
nuclear matter into (nucleons and) several fragments of size $A_{\rm f}>5$, has
been observed in proton-induced and heavy ion reactions \cite{Hue85,Mor93}.
Very recently, similar multifragmentation events have also been observed
in the disintegration of C$_{60}$-fullerenes (`buckyballs') after bombardment
with high-energy heavy ions \cite{Leb94}.  It is clear by now that there
is probably no one single mechanism responsible for the multitude of
different fragmentation scenarios which range from sequential binary emission
processes at low excitation energies \cite{Fri83} over statistical
multifragmentation \cite{Bon82} to explosive vaporization.
The probably most exciting possibility is the occurrence of a
2$^{\rm nd}$-order phase transition at the critical point of nuclear matter
\cite{Min82,Bau85,Cam86}.

The majority of theoretical investigations has been aimed at understanding
the fragment mass distributions.  However, there is considerable
interest in fragment kinetic energy spectra, and the present letter is also
concerned with this topic.  In high-energy proton-induced \cite{Min82} and in
relativistic projectile fragmentation \cite{Gre75} reactions one observes
Boltzmann-like kinetic energy spectra of intermediate mass fragments with
temperatures, for which the fragment mass dependence has been parameterized
\cite{Min82} as $T_{\rm f} = T_o~(A_{\rm r} - A_{\rm f})/A_{\rm r}$,
where $A_{\rm r}$ is the mass of the recoil residue.  The temperature $T_o$
typically has a value of $\approx$ 15 MeV, much larger that nuclear
temperatures extracted from isotope ratios or level population
ratios \cite{Fri94}, and also
much larger than typical temperatures ($\approx 5-8$ MeV)
used in statistical models \cite{Bon82} used to
reproduced the experimental mass yield curves.

Here it is argued that the comparatively high value of $T_{\rm f}$
is a consequence of the
addition of the Fermi momenta of the individual nucleons in the fragment and
thus a consequence of the Fermi-Dirac nature of nucleons.  The
single-particle model is employed.  It previously was successfully used
\cite{Fes73,Gol74,Ber81} to explain the observed dispersion in fragment
transverse momentum spectra generated in projectile fragmentation.

One starts with an ensemble of $A$ nucleons at (internal) temperature
$T_{\rm in}$. In the single-particle picture their momentum distribution
(in the non-relativistic limit) is given by
\begin{equation}
\label{sp}
  \rho({\bf p}) = (1 + \exp[(p^2/2m-\mu)/T_{\rm in}])^{-1}\ ,
\end{equation}
where $\mu$ is the chemical potential.
Adding up the individual momentum vectors of all $A_{\rm f}$ nucleons
in a fragment
one then arrives at the momentum distribution of the fragment
\begin{equation}
\label{exact}
  \rho({\bf P}_{\rm f}) = \int \prod_{i=1}^{A_{\rm f}}
                \{d^3p_i\,\rho({\bf p}_i)\}
                \delta^3({\bf P}_{\rm f} - \sum_{i=1}^{A_{\rm f}}{\bf p}_i)\ .
\end{equation}

This addition procedure is of course nothing else but a random walk in
momentum space and thus in the class of problems first posed by Pearson
\cite{Pea05}.  Since the single-particle probability distribution has
0 average and a fixed variance $\sigma^2$, the central limit theorem of
Gauss applies in the present case, and -- for sufficiently large number of
steps, $A_{\rm f}$, in the random walk -- the probability distribution
can be written as
\begin{equation}
  \rho({\bf P}_{\rm f}) = \frac{1}{\sqrt{2\pi A_{\rm f}\sigma^2}}
           \exp\left(-\frac{P_{\rm f}^2}{2A_{\rm f}\sigma^2}\right)\ .
\end{equation}

Using the non-relativistic approximation
$E_{\rm f}=P_{\rm f}^2/(2m_N A_{\rm f})$,
we therefore find an exponentially falling kinetic energy spectrum for the
fragments
\begin{equation}
\label{gauss}
  \rho(E_{\rm f}) = \frac{2}{\sqrt{\pi\,T_{\rm f}^3}} \sqrt{E_{\rm f}}
      \exp\left(-\frac{E_{\rm f}}{T_{\rm f}}\right)\ ,
\end{equation}
where in this picture the apparent temperature, $T_{\rm f}$, is given by
\begin{equation}
\label{tf}
  T_{\rm f} = \sigma^2 / m_N\ .
\end{equation}

What exactly constitutes a sufficiently large number of steps is not {\it a
priory} clear.
However, in Fig.~1 I answer this question numerically by solving
Eq.~\ref{exact} with a dispersion of
$\sigma^2=(2/5)E_F\,m_N = 0.014$ GeV$^2$. (This choice will become more
apparent below).
For each fragment mass $5\times 10^6$ events were generated.  Displayed are
the resulting distributions (histograms) for $A_{\rm f}=2,3,5,12,100$.  One can
clearly see that
even a very small number of steps (between 5 and 12) are sufficient to
approach the limit given by Gauss, Eq.\ \ref{gauss} (solid line).

In the simple Fermi-gas model at 0 temperature, the variance in the
momentum distribution is given by $\sigma^2$ = $p_F^2/5$, where $p_F$ is
the Fermi momentum.  However, as pointed
out by Goldhaber \cite{Gol74}, there is the additional constraint that
the momenta of all $A$ nucleons combined add up to 0.  This leads to
\begin{equation}
\label{sig2}
  \sigma^2 = \frac{A-A_{\rm f}}{A-1}\langle p_i^2\rangle\ .
\end{equation}
The term $(A-Af)/(A-1)$ is the recoil correction due to total momentum
conservation.  It explains the experimentally observed dependence of
$T_{\rm f}$ on $A_{\rm f}$ \cite{Min82}.

Inserting Eq.~\ref{sig2} into Eq.~\ref{tf} leads
to an apparent fragment kinetic temperature
\begin{equation}
\label{t_of_t}
   T_{\rm f}  = \frac{1}{m_N}\frac{A-A_{\rm f}}{A-1}\langle p_i^2\rangle
              = \frac{A-A_{\rm f}}{A-1}\,\frac{2}{3}\langle E_k\rangle\ ,
\end{equation}
where $\langle E_k\rangle$ is the average kinetic energy
per nucleon as obtained
from the single particle model.  In the
finite temperature case considered here this average is
\begin{equation}
\label{e_of_t}
  \langle E_k \rangle = \int_0^\infty d\epsilon
            \frac{\epsilon^{3/2}}{1 + \exp[(\epsilon-\mu)/T_{\rm in}]} \left/
                       \int_0^\infty d\epsilon
            \frac{\epsilon^{1/2}}{1 + \exp[(\epsilon-\mu)/T_{\rm in}]}\right.\
{}.
\end{equation}

In the limit $T_{\rm in}\rightarrow 0$, we obtain $\mu\rightarrow E_F$ and
$(1 + \exp[(\epsilon-\mu)/T_{\rm in}])^{-1}\rightarrow\theta(E_F-\epsilon)$,
and the average of Eq.~\ref{e_of_t} becomes simply
$\langle E_k(T_{\rm in}\!=\!0) \rangle$ = $\frac{3}{5}E_F$, and
consequently, the apparent fragment kinetic temperature at $T_{\rm in}\!=\!0$
will {\em not} be 0, but rather
\begin{equation}
   T_{\rm f}(T_{\rm in}\!=\!0) =
      \frac{A-A_{\rm f}}{A-1}\,\frac{2}{5}E_F\ .
\end{equation}

In the case $T_{\rm in}>0$, the single particle model is still applicable.
Its assumptions are not dependent on any sudden approximation (as frequently
applied in the study of projectile fragmentation), but also hold for a system
in equilibrium at some finite temperature.  This was already pointed out by
Goldhaber \cite{Gol74}.
In this case, the numerical solution of Eq.~\ref{e_of_t} inserted
into Eq.~\ref{t_of_t} will yield the desired answer, the dependence of
the apparent temperature as extracted from fragment kinetic energy spectra
on the `real' temperature of the system.  For small temperatures, $T_{\rm in}$,
on can perform a Sommerfeld expansion of these equations around
$T_{\rm in}=0$.  This yields the approximation
\begin{equation}
  \mu \approx E_F \left(1 - \frac{\pi^2}{12}
              \left(\frac{T_{\rm in}}{E_F}\right)^2
                         + {\cal O}\left(\frac{T_{\rm in}}{E_F}\right)^4\right)
\end{equation}
for the chemical potential and for the apparent fragment temperature
the useful expression
\begin{equation}
\label{ana}
   T_{\rm f}(T_{\rm in}) \approx
      \frac{A-A_{\rm f}}{A-1}\,\frac{2}{5}E_F
          \left(1 + \frac{5\pi^2}{12}\left(\frac{T_{\rm in}}{E_F}\right)^2
                      + {\cal O}\left(\frac{T_{\rm in}}{E_F}\right)^4\right)\ ,
\end{equation}
which works to better than five percent up to $T_{\rm in}/E_F$ = 0.3.
Fig.~2 shows a comparison of the numerical solution (ignoring the recoil
correction term $(A-A_{\rm f})/(A-1)$) (solid line) as well as
the approximation of Eq.~\ref{ana}.

Bertsch \cite{Ber81} has pointed out that there is (for $T_{\rm in}=0$)
a suppression of the dispersion due to Pauli correlations, which can amount
to about 30\% in the projectile fragmentation of $^{40}$Ca.  This effect is not
included in the simple estimate provided here.

The effects of final state interaction, the emission of a nucleon from the
excited pre-fragment, on the fragment kinetic energy distribution is also
small \cite{Ber81}.  Typical thermal momenta of single nucleons are
$p_t \approx \sqrt{3\,m_n\,T_{\rm in}}$.  This results in corrections
of the order $(p_t/P_{\rm f})^2 \approx T_{\rm in}/A_{\rm f} T_{\rm f}$,
about 5\%
for fragments like carbon.  Final state Coulomb interaction will mainly shift
the energies upwards, and with it raise the temperature $T_{\rm f}$ slightly.

The exact value of $T_{\rm f}$
as a function of $T_{\rm in}$ depends on $E_F$, which in turn depends on the
freeze-out density as $E_F(\rho) = E_F(\rho\!=\!\rho_0)\,(\rho/\rho_0)^{2/3}$.
It is, perhaps, instructive to insert some typical numbers:  If we assume a
freeze-out density of $0.5\,\rho_0$, and a temperature of $T_{\rm in}=6$ MeV,
then we obtain a Fermi energy of 24 MeV, and a slope constant (ignoring the
recoil correction) of $T_{\rm f}=12$ MeV.  This number is of the order
obtained by projectile fragmentation reactions, and it therefore is possible
that this kind of fragmentation reactions are not really a cold breakup of
nuclei at nuclear matter density (as is conventionally assumed), but the
fragmentation of a more dilute system at higher internal temperature.

In summary, by application of the single particle model it was shown that it is
possible to understand the high values of apparent fragment kinetic energy
spectra temperatures as compared to temperatures extracted from isotope ratios
or level population ratios in proton-induced or reverse kinematics
fragmentation reactions.
One may speculate that this basic mechanism can also qualitatively
explain similar observations
for heavy ion induced fragmentation events with sizeable compression (by
inclusion of radial flow momentum components), but the
model employed here might be too simple for this case.

%--------------

This work was supported by the National Science Foundation under Grant
No.~PHY-9403666 and a US National Science Foundation Presidential Faculty
Fellow award.  In addition, I thank the Institute for Nuclear Theory at the
University of Washington for its hospitality and the US Department of Energy
for partial support during the completion of this work.  Many useful
discussions with the participants in the workshop `Hot and Dense Nuclear
Matter', in particular J. Bondorf, G. Bertsch, J. Randrup, and W. Friedman,
are acknowledged.

\clearpage

\section*{Figure Captions}

\begin{description}

\item[Fig.~1] Fragment kinetic energy spectra for different masses (histograms)
      from solving
      Eq.\ \ref{exact} with a value for the variance of 0.014 GeV$^2$. The
      solid line is the Gauss limit, Eq.\ \ref{gauss}.

\item[Fig.~2] Apparent temperature of fragment kinetic energy spectra (in units
      of the Fermi energy) as a function of the temperature, $T_{\rm in}$ of
      the Fermi gas.  Solid line:
      numerical solution of Eq.\ \ref{e_of_t} inserted
      into Eq.\ \ref{t_of_t}.
      Dashed line: analytic approximation, Eq.\ \ref{ana}.

\end{description}
\end {document}